\documentclass[aps,pra,amsmath,amssymb,amsfonts,showpacs]{revtex4}
\usepackage{graphicx}
\def\ket#1{|#1\rangle}
\def\bra#1{\langle #1|}

\def\bracketi#1#2{\langle #1|#2 \rangle}

\def\ketbra#1#2{| #1 \rangle \langle #2 |}
\begin{document}
\title{ Nonlinear interaction of two photons at a one-dimensional atom: spatiotemporal quantum coherence in the emitted field}
\author{Kunihiro Kojima$^{1}$}
\email{kuni@es.hokudai.ac.jp}
\author{Holger F. Hofmann$^{1,2}$}
\author{Shigeki Takeuchi$^{1,2}$}
\author{Keiji Sasaki$^{1}$}
\affiliation{$^{1}$Research Institute for Electronic Science, Hokkaido University, Kita-12 Nishi-6, Kita-ku, Sapporo 060-0812, Japan \\ $^{2}$PRESTO, Japan Science and Technology Corporation (JST), Hokkaido University, Kita-12 Nishi-6, Kita-ku, Sapporo 060-0812, Japan}
\begin{abstract}
The nonlinear photon-photon interaction mediated by a single two-level atom is studied theoretically based on a one-dimensional model of the field-atom interaction. This model allows us to determine the effects of an atomic nonlinearity on the spatiotemporal coherence of a two photon state. Specifically, the complete two photon output wave function can be obtained for any two photon input wave function. It is shown that the quantum interference between the components of the output state associated with different interaction processes causes bunching and anti-bunching in the two photon statistics. This theory may be useful for various applications in photon manipulation, e.g. quantum information processing using photonic qubits, quantum nondemolition measurements, and the generation of entangled photons.
\end{abstract}
\maketitle
\section{\label{sec:level1}Introduction}
The nonlinearity of atomic objects, e.g. two-level atoms and quantum dots, can be sensitive to individual photons. This kind of nonlinearity may be useful for the study of photon manipulations such as quantum information processing \cite{fredkin,turchette,holkoji,nielsen}, quantum nondemolition measurements \cite{nondemo}, and the generation of entangled photons \cite{entsource}. Realizations of single atom nonlinearities have been studied extensively in the field of cavity quantum electrodynamics \cite{rice,berman,turchetteb}. The sensitivity of this atom-cavity system to individual photons has been demonstrated by Turchette et al. \cite{turchette}. Recently, we proposed a modification of this setup that enhances the nonlinearity by avoiding all losses in a one-sided cavity geometry \cite{holkoji}. Every input photon will then be found in the output. The nonlinear response of a single atom to an input of two photons, e.g. from single photon sources \cite{yamamoto}, can then be observed in the correlations between the two output photons. In order to apply the correlations due to the nonlinear photon-photon interaction, e.g. in quantum information processing, it is important to understand the precise spatiotemporal coherence of the input and output photons. This cannot be fully achieved by theories that eliminate the quantum state of the field outside the atom-cavity system \cite{rice,berman,turchetteb}. We therefore propose a theory that includes the propagation to and from the system in the quantum state, based on a one-dimensional model of the field-atom interaction \cite{holger2}. In this paper, we apply this one-dimensional model of the field-atom interaction to the case of two photon input wave packets.

The rest of this paper is organized as follows. In section \ref{sec:level2}, we give a theoretical model of the light-atom interaction in one-dimensional free space.  In section \ref{sec:level3}, we discuss the experimental realization. In section \ref{sec:level4}, we derive the general solution of the Schr\"odinger equation for the one photon case. In section \ref{sec:level5}, we apply the result of the one photon case to derive the general solution for two photons. In section \ref{sec:level6}, the analysis of spatiotemporal coherence in the outgoing wave packet is performed as an example for the applications of our theory. In section \ref{sec:level7}, the effects of the nonlinear interaction on the second order correlations is discussed. In section \ref{sec:level8}, it is shown that the bunching and anti-bunching effects in the two photon statistics can be understood as quantum interference effects of different output components. In  section \ref{sec:level9}, we conclude with a summary of our discussions.
\section{\label{sec:level2}Theoretical model of the light-atom interaction in one-dimensional free space}
\begin{figure}[ht]
\begin{center} 
\includegraphics[width=5cm,]{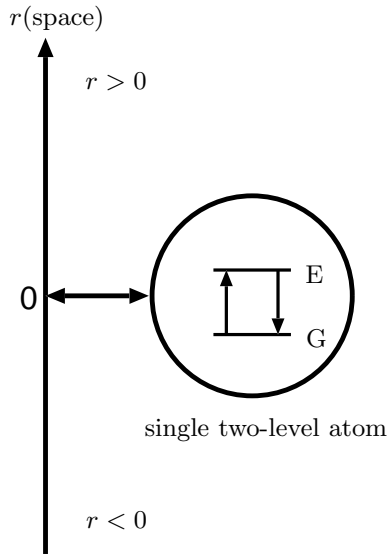}
\caption{\label{fig:figure1}Illustration of the theoretical model. The r-axis represents the single spatial coordinate of the field. A single two-level atom is placed at the position \(r=0\). {\rm E} and {\rm G} represent the ground state and the excited state of the atom. \(r > 0\) corresponds to the output field and \(r < 0\) corresponds to the input field.}
 \end{center} 
\end{figure}

To investigate the change of the spatiotemporal quantum coherence originating from the nonlinear interaction of two photons mediated by a single two-level system, we need a model for the spatiotemporal propagation to and from the atom. A possible model has been studied in the analysis of spatiotemporal quantum coherence for the case of spontaneous emission from a single excited atom \cite{holger2}. This model is illustrated in fig.~\ref{fig:figure1}. The r-axis represents the single spatial coordinate of the one-dimensional light field. A single two-level atom is coupled locally with the light field at the position \(r=0\).  The negative region \(r < 0\) and the positive region \(r > 0\) correspond to the incoming field and the outgoing field, respectively. This means that the light field can only propagate in the positive direction, approaching the atom at \(r < 0\), and moving away from it at \(r > 0\). The dispersion relation describing the field dynamics is then given by the wave number multiplied by the speed of light, \(\omega = c k\).
 The Hamiltonian composed of the uninterrupted propagation and the interaction between the atom and the one-dimensional field can be written as
\begin{eqnarray}
\hat{H} &=& \int^{\infty}_{-\infty} dk \ \hbar c k \; \hat{b}^{\dagger}(k) \hat{b}(k) + \int^{\infty}_{-\infty}dk \ i\hbar \sqrt{\frac{c \Gamma}{\pi}} \left( \hat{b}^{\dagger}(k) \hat{\sigma}_{-}-\hat{\sigma}^{\dagger}_{-}\hat{b}(k) \right), \label{eq:hamiltonian}
\end{eqnarray}
where \(\hat{b}(k)\) is the photon annihilation operator, and \(\sigma_{-}\) is the annihilation operator of the atomic excitation. \(\sqrt{c \Gamma /\pi}\) is the coupling constant between the atom and the light field. \(\Gamma\) is the dipole relaxation rate. As will be seen later, this rate defines the only relevant time scale of our model. Note that the Hamiltonian has been formulated in a rotating frame defined by the transition frequency \(\omega_{0}\). Likewise, the wave vector \(k\) is defined in the rotating frame, i.e. \(k\) is defined relative to the resonant wave vector \(\omega_{0}/c\).
\section{\label{sec:level3}Experimental realization}
\begin{figure}[ht]
\begin{center} 
\includegraphics[width=8cm]{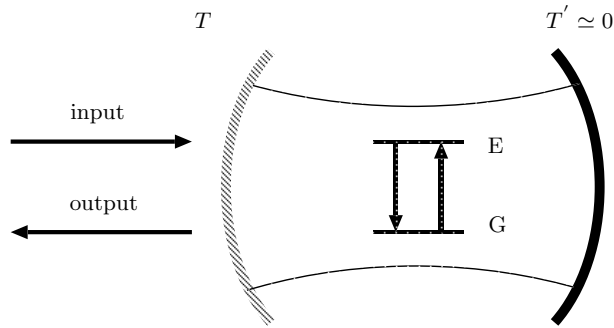}
\caption{\label{fig:figure2}Schematic representation of cavity geometry. \(T\) and \(T^{`}\) are the transmittances of the mirrors. {\rm E} and {\rm G} represent the ground state and the excited state of the single two-level atom. The arrows to the left of the cavity represent the free space input and output fields, respectively.}
 \end{center} 
\end{figure}

The situation described by the theoretical model in section \ref{sec:level2} can be realized  experimentally by using a one-sided cavity as illustrated in fig.~\ref{fig:figure2}. The left mirror of the cavity has a transmittance much higher than the right mirror, which has nearly 100 \% reflectance. The negative region on the space axis in the model shown in fig.~\ref{fig:figure1} corresponds to the input in fig.~\ref{fig:figure2} and the positive region corresponds to the output in fig.~\ref{fig:figure2}.

In terms of the conventional cavity quantum electrodynamics parameters, this regime is characterized by \(\kappa \gg g\), where \(\kappa\) is the cavity damping rate through the left mirror and \(g\) is the dipole coupling between the atom and the cavity mode. Therefore the method of adiabatic elimination can be applied to the time evolution of the cavity field \cite{rice}. This means that since the cavity damping rate \(\kappa\) is much faster than the dipole coupling \(g\), the interaction between the atom and the outside field mediated by the cavity field can be expressed by an effective dipole relaxation rate \(\Gamma = g^{2}/\kappa\). It should be noted that, in this case, the Hamiltonian (1) represents an approximation valid only within the finite cavity bandwidth of $2\kappa$. Effectively, the theoretical model can then be used to correctly describe the atom-cavity dynamics at timescales larger than $1/\kappa$. This approximation is sufficient for the description of atomic absorption and emission processes as long as $1/\Gamma \gg 1/\kappa$. Features of the cavity response which become obvious only at a time scale of about $1/\kappa$  are neglected in this paper.

The dipole relaxation rate $\Gamma$ describes the dipole damping caused by emissions through the left mirror of the cavity, and the corresponding rate of spontaneous emission through the cavity is equal to \(2\Gamma\) \cite{rice,turchetteb,holkoji}. In our case, we assume that the rate of spontaneous emission into the non-cavity modes \(\gamma_{\|}\) is negligible (\(\gamma_{\|} \ll 2\Gamma\)). Nearly all emissions from the atom can then be confined to the cavity and 2\(\Gamma\) is the total spontaneous emission rate of the excited atom in the cavity. In present cavity designs, this can be realized by covering a large solid angle of the atomic emission with the cavity mirrors and exploiting the enhancement of spontaneous emission by the cavity. For example, in the case of Turchette et al.'s experiment \cite{turchette}, the cavity parameters indicate that about 70\% of the sponaneous emission from the atom is emitted through the single cavity mode. Another promising method of achieving a one-dimensional emission and absorption of the atom is the use of semiconductor microstructures, as reported e.g. in \cite{solomon}. In any case, our model should apply to any cavity design with \(\kappa \gg g \gg \gamma_{\|}\).
\section{\label{sec:level4}One photon processes}
In this section, we treat the interaction of one photon with the atomic system as a preparation for the analysis of two photon processes. We can expand the quantum state of the single photon in the basis of the wave number eigenstates \(\ket{k}\) and the excited state \(\ket{\text{\rm E}}\) of the two-level atom. The quantum state for the one photon process can then be written as
\begin{equation}
\ket{\Psi(t)} = \Psi({\rm E};t) \ket{{\rm E}} + \int^{\infty}_{-\infty} dk \ \Psi(k;t) \ket{k}. \label{eq:onephoton}
\end{equation}
In this basis, The Hamiltonian given by (\ref{eq:hamiltonian}) can be expressed as
\begin{eqnarray}
\hat{H}_{\text{1photon}} &=& \hbar c \hat{k} + \hat{H}_{\text{int}} \label{eq:hamiltonian1}\\
\text{with} && {} \hat{k} = \int^{\infty}_{-\infty} dk\ k \ketbra{k}{k}\ \ \text{and} \ \ \hat{H}_{\text{int}} = i\hbar \sqrt{\frac{c\Gamma}{\pi}} \int^{\infty}_{-\infty} dk\ \left( \ketbra{k}{{\rm E}}-\ketbra{{\rm E}}{k} \right). \nonumber\end{eqnarray} 
The equations for the time evolution of the probability amplitudes \(\Psi({\rm E};t)\) and \(\Psi(k;t)\) can then be obtained from the Schr\"odinger equation \(i\hbar\ d\!/\!dt\ \ket{\Psi(t)} = \hat{H} \ket{\Psi(t)}\) using (\ref{eq:onephoton}) and (\ref{eq:hamiltonian1}),
\begin{eqnarray}
\frac{d}{dt} \Psi({\rm E};t) &=& - \sqrt{\frac{c \Gamma}{\pi}} \int^{\infty}_{-\infty} dk \ \Psi(k;t) \label{eq:absorption}\\
\frac{d}{dt}\Psi(k;t) &=& -i k c \Psi(k;t) + \sqrt{\frac{c \Gamma}{\pi}} \Psi({\rm E};t). \label{eq:emittion}
\end{eqnarray}
The time evolution \(\Psi(k;t)\) can be obtained by integrating (\ref{eq:emittion}),
\begin{equation}
\Psi(k;t) = e^{-ikc (t-t_{i})} \Psi(k;t_{i}) + \sqrt{\frac{c \Gamma}{\pi}} \int^{t}_{t_{i}} dt^{'} \ e^{-ikc (t-t^{'})} \Psi({\rm E};t^{'}), \label{eq:emitsolution}
\end{equation}
where \(t_{i}\) is the initial time of the time evolution. In order to describe the time evolution in real space, the result of the integration (\ref{eq:emitsolution}) can be Fourier transformed using
\begin{equation}
\Psi(r;t) =\frac{1}{\sqrt{2\pi}} \int^{\infty}_{-\infty} dk \ e^{ikr} \Psi(k,t). \label{eq:fourietransform}
\end{equation}
The real space representation of the time evolution then reads
\begin{equation}
\Psi(r;t) = 
\begin{cases}
\Psi(r-c(t-t_{i});t_{i}) & \text{for $r < 0$ or $c(t-t_{i}) < r$}\\
\Psi(r-c(t-t_{i});t_{i}) + \sqrt{\frac{2 \Gamma}{c}} \Psi({\rm E};t-\frac{r}{c}) & \text{for $0 < r < c(t-t_{i})$.}
\end{cases} \label{eq:onephotonsol}
\end{equation}
The top term corresponds to the single photon amplitude propagating without being absorbed by the atom. On the other hand, the bottom term consists of two processes. The first term also corresponds to propagation without absorption, while the second term corresponds to the amplitude of a single photon reemitted into the outgoing field after absorption by the atom \cite{holger2}. The time evolution \(\Psi({\rm E};t)\) of the excited state amplitude can be obtained by integrating (\ref{eq:absorption}) using the result for \(\Psi(k;t)\) given in (\ref{eq:emitsolution}),
\begin{eqnarray}
\Psi({\rm E};t) &=& - \sqrt{\frac{c \Gamma}{\pi}} \int^{t}_{t_{i}} dt \int^{\infty}_{-\infty} dk \ \Psi(k;t) \nonumber\\
&=& - \sqrt{\frac{c \Gamma}{\pi}} \int^{t}_{t_{i}} dt \int^{\infty}_{-\infty} dk \ \left( e^{-ikc (t-t_{i})} \Psi(k;t_{i}) + \sqrt{\frac{c \Gamma}{\pi}} \int^{t}_{t_{i}} dt^{'} \ e^{-ikc (t-t^{'})} \Psi({\rm E};t^{'}) \right)
\end{eqnarray}
Using the Fourier transform to obtain the real space representation of \(\Psi(k;t)\), the result reads
\begin{eqnarray}
\Psi({\rm E};t) = e^{- \Gamma (t-t_{i})}\Psi({\rm E};t_{i})- \sqrt{2 \Gamma c} \int^{t}_{t_{i}} dt^{'} \ e^{-\Gamma (t-t^{'})} \Psi(-c(t^{'}-t_{i});t_{i}). \label{eq:absorptionfunction}
\end{eqnarray}
The first term corresponds to emission from the excited atom and the second term corresponds to the excitation of the atom by the incoming light.

Since we are interested in the response of the ground state atom to a one photon input, the initial conditions of \(\Psi({\rm E};t_{i})\) and \(\Psi(r;t_{i})\) can be defined as follows,
\begin{eqnarray}
\Psi(r > 0;t_{i}) &=& 0, \label{eq:condition1}\\
\Psi({\rm E};t_{i}) &=& 0. \label{eq:condition2}
\end{eqnarray}
Condition (\ref{eq:condition1}) represents the assumption that there is no light at \(r > 0\). Condition (\ref{eq:condition2}) represents the assumption that the atom is initially in the ground state. With the above conditions, the time evolution \(\Psi(r;t)\) given by (\ref{eq:onephotonsol}) and (\ref{eq:absorptionfunction}) becomes
\begin{equation}
\Psi(r;t) = 
\begin{cases}
\Psi(r-c(t-t_{i});t_{i}) & \text{for $r < 0$}\\
\Psi(r-c(t-t_{i});t_{i}) - 2\Gamma \int^{t-r/c}_{t_{i}} dt^{'} \ e^{-\Gamma (t-r/c-t^{'})} \Psi(-c(t^{'}-t_{i});t_{i})  & \text{for $0 < r < c(t-t_{i})$}\\
0 & \text{for $r > c(t-t_{i})$}.
\end{cases} \label{eq:onephotonsol2}
\end{equation}

\begin{figure}[ht]
\begin{picture}(0,0)
\put(-15,170){(a)}
\put(-15,-22){(b)}
\end{picture}
\includegraphics[width=10cm]{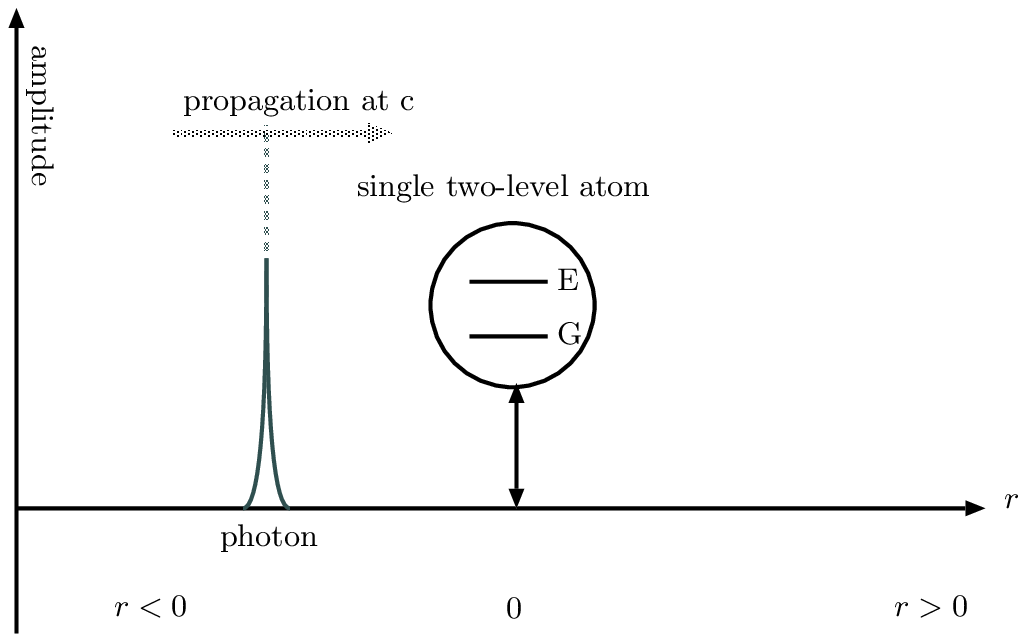}\\
\vspace*{0.5cm}
\includegraphics[width=10cm]{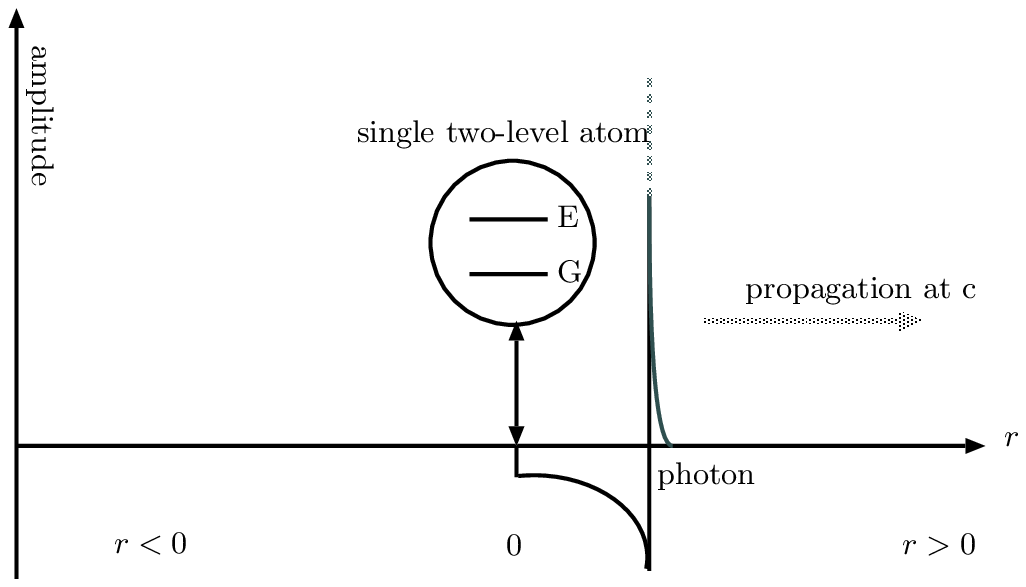}
\caption{\label{fig:figure3}Schematic representation of the time evolution of incoming and outgoing amplitude. The horizontal axes in (a) and (b) represent the space coordinate \(r\). The vertical axes represent the probability amplitude of the spatial one-dimensional field. (a) illustrates the amplitude before the arrival of the photon at the atom and (b) illustrates the amplitude after the arrival of the photon at the atom. In (a), the photon, having a delta like amplitude, is propagating in the incoming field. The amplitude at time t has its peak at \(r=c(t-t_{i})+r^{'}\). In (b), the outgoing amplitude of the photon is in a superposition of two amplitudes associated with the uninterrupted propagation of the photon and with reemission after absorption by the atom, respectively.}
\end{figure}

To investigate the outgoing amplitude \(\Psi(r > 0;t)\) for an arbitrary incoming amplitude under the conditions (\ref{eq:condition1}) and (\ref{eq:condition2}), it is convenient to represent \(\Psi(r > 0;t)\) by using the matrix element of the time evolution operator. In order to derive this matrix element, we expand the time evolution \(\ket{\Psi(t)}\) of the quantum state given by (\ref{eq:condition1}) and (\ref{eq:condition2}) as follows,
\begin{eqnarray}
\ket{\Psi(t)} &=& \hat{U}(t-t_{i}) \ket{\Psi(t_{i})} \nonumber\\
              &=& \int^{\infty}_{-\infty} drdr^{'} \ket{r} \bra{r} \hat{U}(t-t_{i}) \ket{r^{'}} \bracketi{r^{'}}{\Psi(t_{i})} \nonumber\\
              &+& \ket{{\rm E}} \int^{\infty}_{-\infty} dr^{'} \bra{{\rm E}} \hat{U}(t-t_{i}) \ket{r^{'}} \bracketi{r^{'}}{\Psi(t_{i})}, \label{eq:ketexpand}
\end{eqnarray}
where \(\ket{r} \equiv \frac{1}{\sqrt{2\pi}}\int^{\infty}_{-\infty} dk \ e^{-ikr} \ket{k}\) is the eigenstate of a photon at the position \(r\). \(\hat{U}(t-t_{i}) = e^{-\frac{i}{\hbar} \hat{H}_{\text{1photon}}(t-t_{i})}\) is the time evolution operator. \(\Psi(r;t)\) can then be expressed as follows,  
\begin{eqnarray}
\Psi(r;t) &=& \bracketi{r}{\Psi(t)} \nonumber \\
              &=& \int^{\infty}_{-\infty} dr^{'} {\bf u}_{\text{1photon}}(r,r^{'};t-t_{i}) \Psi(r^{'};t_{i}),\label{eq:expandpsi}\\
\text{where} && {} {\bf u}_{\text{1photon}}(r;r^{'};t-t_{i}) = \bra{r} \hat{U}(t-t_{i}) \ket{r^{'}}. \nonumber
\end{eqnarray}
\({\bf u}_{\text{1photon}}(r;r^{'};t-t_{i})\) is the matrix element of the time evolution operator \(\hat{U}(t-t_{i})\). This matrix element is the transition probability amplitude from the state \(\ket{r^{'}}\) at time \(t_{i}\) to the state \(\ket{r}\) at time \(t\). \({\bf u}_{\text{1photon}}(r,r^{'};t-t_{i})\) can be obtained by comparing the results (\ref{eq:onephotonsol2}) and (\ref{eq:expandpsi}). It is given by 
\begin{eqnarray}
{\bf u}_{\text{1photon}}(r;r^{'};t-t_{i}) &=& {\bf u}_{\text{prop}}(r;r^{'};t-t_{i}) + {\bf u}_{\text{abs}}(r;r^{'};t-t_{i}) \nonumber \\
\text{with} && {} {\bf u}_{\text{prop}}(r;r^{'};t-t_{i}) = \delta(r-c(t-t_{i})-r^{'}) \nonumber\\
\text{and} && {\bf u}_{\text{abs}}(r;r^{'};t-t_{i}) = 
\begin{cases}
-\frac{2\Gamma}{c} e^{-\frac{\Gamma}{c}(c(t-t_{i}) + r^{'}-r)} & \text{for $0 < r < c(t-t_{i})+r^{'}$ and $r^{'} < 0$}\\
0 & \text{for $r > c(t-t_{i})+r^{'}$ or $r^{'} >0$}.
\end{cases} \label{eq:onephotonprocesses} 
\end{eqnarray}
This transition amplitude can be interpreted as follows. At the initial time \(t_{i}\), the photon starts to propagate at the position \(r^{'}\). For \( r < 0 \), it propagates at constant velocity \(c\). For \(r > 0\), it has already passed the atom and it is in a superposition of two amplitudes corresponding to the uninterrupted propagation of the photon and to absorption and reemission by the atom, respectively. \({\bf u}_{\text{prop}}\) corresponds to the uninterrupted propagation and \({\bf u}_{\text{abs}}\) corresponds to the reemission from the atom at an emission rate of \(\Gamma\). An illustration of the one photon time evolution is shown in fig.~\ref{fig:figure3}. 
\section{\label{sec:level5}Two photon processes}
In this section, we treat the interaction of two photons with a single two-level atom. The method we use here to describe the two photon quantum state is to first distinguish the particles and then to introduce the correct symmetry of the wavefunction for indistinguishable bosons. Note that this is a standard textbook approach to problems in multi particle quantum mechanics \cite{dirac}. If we consider the two photons to be separate physical systems, we can describe their quantum state in the product space of their single photon Hilbert spaces. A general state \(\ket{\Psi(t)}\) for two photon processes can then be expanded in the product basis of the eigenstates \(\ket{k}\) and \(\ket{\rm E}\) as
\begin{eqnarray}
\ket{\Psi(t)} &=& \int^{\infty}_{-\infty} dk_{1} \ \Psi(k_{1};{\rm E}_{2};t) \ket{k_{1}} \otimes \ket{{\rm E}_{2}} + \int^{\infty}_{-\infty} dk_{2} \ \Psi({\rm E}_{1};k_{2};t) \ket{{\rm E}_{1}} \otimes \ket{k_{2}} \nonumber\\
&+& \int^{\infty}_{-\infty} dk_{1} dk_{2} \ \Psi(k_{1};k_{2};t) \ket{k_{1}} \otimes \ket{k_{2}} + \Psi({\rm E}_{1};{\rm E}_{2};t) \ket{{\rm E}_{1}} \otimes \ket{{\rm E}_{2}}, \label{eq:state}
\end{eqnarray}
where the indices \(1\) and \(2\) of the eigenstates \(\ket{k}\) and \(\ket{\rm E}\) distinguish the two photons. Since photons are indistinguishable bosons, the actual quantum state has to be a state of positive symmetry. This bosonic property of photons requires that the quantum states fulfill the following conditions,
\begin{eqnarray}
\Psi(k;{\rm E}_{2};t) &=& \Psi({\rm E}_{1};k;t) \nonumber \\ 
\Psi(k_{1};k_{2};t) &=& \Psi(k_{2};k_{1};t).
\end{eqnarray}

In order to make it easier to formulate the matrix element of the time evolution for two photon processes, we will first divide the Hamiltonian into a linear term and a nonlinear term. The linear term describes the dynamics of two photons that are absorbed and emitted independently. The Hamiltonian of this interaction free dynamics is then given by the sum of the single photon Hamiltonians. Note that this linear Hamiltonian includes transitions to the state \(\ket{{\rm E}_{1},{\rm E}_{2}}\), where both photons are absorbed by the atom. Since this transition is impossible in a two-level atom, there will be an interaction between the two photons. This interaction can be described by a nonlinear term that suppresses the transitions to the state \(\ket{{\rm E}_{1},{\rm E}_{2}}\). In the following, we first introduce the linear term and then solve the Schr\"odinger equation for the linear system using the results of the one photon case. Next, the nonlinear term is added to the linear term and the matrix element of the time evolution for two photon processes is derived by a comparison between the linear equations and the nonlinear equations.

The linear component of the two photon Hamiltonian can be expressed using the single photon Hamiltonian (\ref{eq:hamiltonian1}),
\begin{eqnarray}
\hat{H}^{\bf{lin}} &=& \hat{H}^{(1)}_{\text{1photon}} \otimes \hat{I}^{(2)}_{\text{1photon}} + \hat{I}^{(1)}_{\text{1photon}} \otimes \hat{H}^{(2)}_{\text{1photon}}, \nonumber\\
\text{where} && {} \hat{H}^{(i)}_{\text{1photon}} = \hbar c \hat{k}^{(i)} + \hat{H}^{(i)}_{\text{int}}\ \ \text{and} \ \ \hat{I}^{(i)}_{\text{1photon}} = \int^{\infty}_{-\infty} dk_{i} \ketbra{k_{i}}{k_{i}} + \ketbra{{\rm E}_{i}}{{\rm E}_{i}} \nonumber \\
\text{with} && {} \hat{k}^{(i)} =\int^{\infty}_{-\infty} dk_{i} \ k_{i} \ketbra{k_{i}}{k_{i}}\ \ \text{and}\ \ \hat{H}^{(i)}_{\text{int}} = i\hbar \sqrt{\frac{c\Gamma}{\pi}} \int^{\infty}_{-\infty} dk_{i} \ \left( \ketbra{k_{i}}{{\rm E}_{i}}-\ketbra{{\rm E}_{i}}{k_{i}} \right). \label{eq:linearhamiltonian}
\end{eqnarray}
This Hamiltonian describes the interaction free evolution of the two photon state. Therefore the time evolution of the two photon probability amplitudes \(\Psi^{{\bf lin}}(k_{1},k_{2};t)\), \(\Psi^{{\bf lin}}({\rm E}_{1},k_{2};t)\) and \(\Psi^{{\bf lin}}(k_{1},{\rm E}_{2};t)\) given by this Hamiltonian corresponds to the results of the one photon case given in (\ref{eq:absorption}-\ref{eq:emittion}). It reads
\begin{eqnarray}
\frac{d}{dt} \Psi^{{\bf lin}}(k_{1},k_{2};t) &=& -ic (k_{1}+k_{2}) \Psi^{{\bf lin}}(k_{1},k_{2};t) + \sqrt{\frac{c\Gamma}{\pi}} \left( \Psi^{{\bf lin}}(k_{1},{\rm E}_{2};t) + \Psi^{{\bf lin}}({\rm E}_{1},k_{2};t)\right) \label{eq:holgerlikelin1}\\
\frac{d}{dt} \Psi^{{\bf lin}}({\rm E}_{1},k_{2};t) &=& -ick_{2} \Psi^{{\bf lin}}({\rm E}_{1},k_{2};t) - \sqrt{\frac{c \Gamma}{\pi}} \int^{\infty}_{-\infty} dk_{1} \ \Psi^{{\bf lin}}(k_{1},k_{2};t) + \sqrt{\frac{c \Gamma}{\pi}} \Psi^{{\bf lin}}({\rm E}_{1},{\rm E}_{2};t) \label{eq:holgerlikelin2}\\
\frac{d}{dt} \Psi^{{\bf lin}}(k_{1},{\rm E}_{2};t) &=& -ick_{1} \Psi^{{\bf lin}}(k_{1},{\rm E}_{2};t) - \sqrt{\frac{c \Gamma}{\pi}} \int^{\infty}_{-\infty} dk_{2} \ \Psi^{{\bf lin}}(k_{1},k_{2};t) + \sqrt{\frac{c \Gamma}{\pi}} \Psi^{{\bf lin}}({\rm E}_{1},{\rm E}_{2};t). \label{eq:holgerlikelin3}
\end{eqnarray}
Since the time evolution described by \(\hat{H}^{{\bf lin}}\) corresponds to the time evolution of the single photon case, the integration can be performed according to the procedure in eqs. (\ref{eq:emitsolution}-\ref{eq:absorptionfunction}), and the matrix element \({\bf u}^{{\bf lin}}_{\text{2photon}}(r_{1},r_{2};r_{1}^{'},r_{2}^{'};t-t_{i})\) of the time evolution operator can be expressed as a product of the individual single photon matrix elements given by (\ref{eq:onephotonprocesses}), 
\begin{eqnarray}
{\bf u}^{{\bf lin}}_{\text{2photon}}(r_{1},r_{2};r^{'}_{1},r^{'}_{2};t-t_{i}) &=& {\bf u}_{\text{1photon}}(r_{1};r_{1}^{'};t-t_{i}) \cdot {\bf u}_{\text{1photon}}(r_{2};r_{2}^{'};t-t_{i}). \label{eq:linearpropagator}
\end{eqnarray}
The photon-photon interaction can now be included by adding the nonlinear term that suppresses transitions to the two photon absorption state \(\ket{{\rm E}_{1},{\rm E}_{2}}\),
\begin{eqnarray}
\hat{H}_{\text{2photon}} &=& \hat{H}^{\bf{lin}} + \Delta \hat{H}^{\bf{Nonlin}},  \nonumber \\
\text{where} && {} \Delta \hat{H}^{\bf{Nonlin}} = -\left( \hat{H}^{(1)}_{\text{int}} \otimes \ketbra{{\rm E}_{2}}{{\rm E}_{2}} + \ketbra{{\rm E}_{1}}{{\rm E}_{1}} \otimes \hat{H}^{(2)}_{\text{int}} \right). \label{eq:nonlinearhamiltonian}
\end{eqnarray}
With this addition, the matrix elements of the two photon Hamiltonian (24)
 are identical with the corresponding two photon matrix elements of the original Hamiltonian (1). The time evolution of the wavefunction is modified only slightly by the addition of the nonlinear term. It now reads
\begin{eqnarray}
\frac{d}{dt} \Psi(k_{1},k_{2};t) &=& -ic (k_{1}+k_{2}) \Psi(k_{1},k_{2};t) + \sqrt{\frac{c\Gamma}{\pi}} \left( \Psi(k_{1},{\rm E}_{2};t) + \Psi({\rm E}_{1},k_{2};t)\right) \label{eq:holgerlike1}\\
\frac{d}{dt} \Psi({\rm E}_{1},k_{2};t) &=& -ick_{2} \Psi({\rm E}_{1},k_{2};t) - \sqrt{\frac{c \Gamma}{\pi}} \int^{\infty}_{-\infty} dk_{1} \ \Psi(k_{1},k_{2};t) \label{eq:holgerlike2}\\
\frac{d}{dt} \Psi(k_{1},{\rm E}_{2};t) &=& -ick_{1} \Psi(k_{1},{\rm E}_{2};t) - \sqrt{\frac{c \Gamma}{\pi}} \int^{\infty}_{-\infty} dk_{2} \ \Psi(k_{1},k_{2};t). \label{eq:holgerlike3}
\end{eqnarray}
The comparison between (\ref{eq:holgerlikelin2}-\ref{eq:holgerlikelin3}) and (\ref{eq:holgerlike2}-\ref{eq:holgerlike3}) shows that the integration of (\ref{eq:holgerlike2}-\ref{eq:holgerlike3}) is the same as the integration of (\ref{eq:holgerlikelin2}-\ref{eq:holgerlikelin3}), except that \(\Psi({\rm E}_{1},{\rm E}_{2};t)\) is zero in the integration of \(\Psi({\rm E}_{1},k_{2};t)\) and \(\Psi(k_{1},{\rm E}_{2};t)\). This means that the matrix element \({\bf u}_{\text{2photon}}(r_{1},r_{2};r^{'}_{1},r^{'}_{2};t-t_{i})\) of the time evolution described by \(\hat{H}_{\text{2photon}}\) can be obtained by identifying the components in the interaction free propagation of the two photons from  \(r^{'}_{1},r^{'}_{2}\) to \(r_{1},r_{2}\) described by the matrix element \({\bf u}^{{\bf lin}}(r_{1},r_{2};r^{'}_{1},r^{'}_{2};t-t_{i})\) and removing the ones due to two photon absorption.

For the case that photon 1 and photon 2 start at \(r^{'}_{2} < r^{'}_{1} < 0\), the time evolution of this probability amplitude can be obtained by distinguishing the following three time regions:
\begin{description}
\item[{\small Time region I}:] \(t < t_{i}-\frac{r^{'}_{1}}{c}\), the two photons are independently propagating in the incoming field (See fig.~\ref{fig:figure4} (a)). The matrix element associated with the uninterrupted propagation \({\bf u}^{{\bf I}}_{\text{2photon}}(r_{1},r_{2};r_{1}^{'},r_{2}^{'};t-t_{i})\) can be expressed as
\begin{equation}
{\bf u}^{{\bf I}}_{\text{2photon}}(r_{1},r_{2};r_{1}^{'},r_{2}^{'};t-t_{i}) = {\bf u}_{\text{prop}}(r_{1};r_{1}^{'};t-t_{i}) \cdot {\bf u}_{\text{prop}}(r_{2};r_{2}^{'};t-t_{i}).
\end{equation}
\item[{\small Time region II}:] \(t_{i}-\frac{r^{'}_{1}}{c} < t < t_{i}-\frac{r^{'}_{2}}{c}\), photon 1 has already arrived at the atom and photon 2 has not arrived yet. In this situation, photon 1 is interacting with the atom and  photon 2 is propagating in the incoming field (See fig.~\ref{fig:figure4} (b)). The matrix element associated with one photon absorption \({\bf u}^{{\bf II}}_{\text{2photon}}(r_{1},r_{2};r_{1}^{'},r_{2}^{'};t-t_{i})\) can be expressed as 
\begin{equation}
{\bf u}^{{\bf II}}_{\text{2photon}}(r_{1},r_{2};r_{1}^{'},r_{2}^{'};t-t_{i}) = {\bf u}_{\text{1photon}}(r_{1};r_{1}^{'};t-t_{i}) \cdot {\bf u}_{\text{prop}}(r_{2};r_{2}^{'};t-t_{i}). 
\end{equation}

\begin{figure}[ht]
\begin{picture}(0,0)
\put(-15,170){(a)}
\put(-15,-22){(b)}
\end{picture}
\includegraphics[width=10cm]{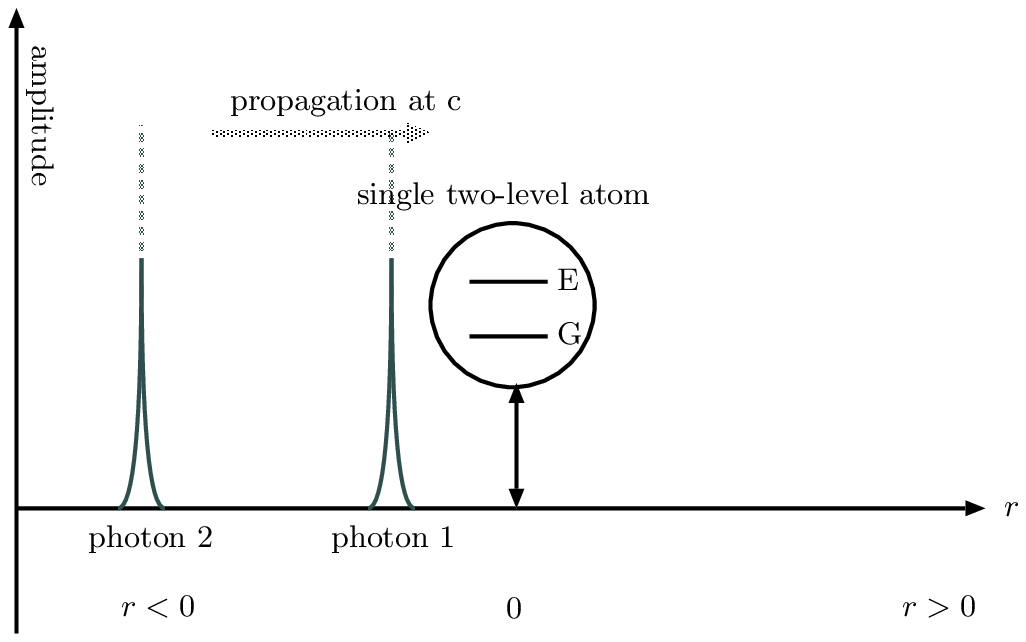}\\
\vspace*{0.5cm}
\includegraphics[width=10cm]{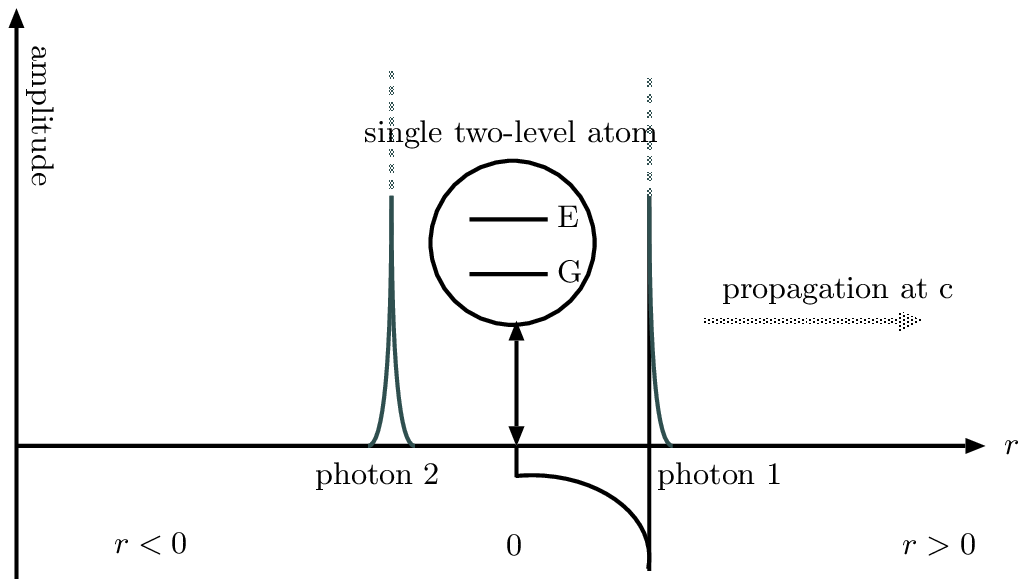}
\caption{\label{fig:figure4}Schematic representation of the time evolution of the incoming and the outgoing amplitude in time region I and II. The horizontal axes in (a) and (b) represent the space coordinate \(r\). The vertical axes represent the probability amplitude of the spatial one-dimensional field. (a) corresponds to time region I, (b) corresponds to time region II. In time region I, photon 1 and photon 2, having a delta like amplitude, are propagating in the incoming field. Their amplitudes at time t have their peaks at \(r_{1}=c(t-t_{i})+r^{'}_{1}\) and \(r_{2}=c(t-t_{i})+r^{'}_{2}\), respectively. In time region II, only photon 1 has arrived at the atom, the outgoing amplitude of photon 1 is as same as the amplitude for the one photon case. Time region III is not shown, since it results in a superposition state that cannot be represented by single photon amplitudes.}
\end{figure}
\item[{\small Time region III}:] \(t_{i}-\frac{r^{'}_{2}}{c} < t\) includes a new situation not treated in the one photon processes, because both photons have now reached the atom. In this time region, the time evolution given by \(\hat{H}_{\text{2photon}}\) is different from the one given by \(\hat{H}^{{\bf lin}}\) because there cannot be any contributions of the two photon absorption amplitude \(\Psi({\rm E}_{1},{\rm E}_{2};t-t_{i})\) in the time evolution of the outgoing amplitude. This means that the matrix element corresponding to two photon absorption \({\bf u}_{\text{abs}}(r_{1};r_{1}^{'};t-t_{i}) \cdot {\bf u}_{\text{abs}}(r_{2};r_{2}^{'};t-t_{i})\) will not be included in the time evolution of \({\bf u}_{\text{2photon}}(r_{1},r_{2};r^{'}_{1},r^{'}_{2};t-t_{i})\) if the position \(r_{1}\) of photon 1 is less than \(c(t-t_{i})+r^{'}_{2}\), since, in this case, photon 2 arrives at the atom before photon 1 is reemitted. The matrix element \({\bf u}^{{\bf III}}_{\text{2photon}}(r_{1},r_{2};r_{1}^{'},r_{2}^{'};t-t_{i})\) can then be expressed as 
\begin{eqnarray}
{\bf u}^{{\bf III}}_{\text{2photon}}(r_{1},r_{2};r_{1}^{'},r_{2}^{'};t-t_{i}) &=& 
\begin{cases}
{\bf u}_{\text{1photon}}(r_{1};r_{1}^{'};t-t_{i}) \cdot {\bf u}_{\text{1photon}}(r_{2};r_{2}^{'};t-t_{i}) & \text{for $r_{1} > c(t-t_{i}) + r^{'}_{2}$}\\
{\bf u}_{\text{1photon}}(r_{1};r_{1}^{'};t-t_{i}) \cdot {\bf u}_{\text{prop}}(r_{2};r_{2}^{'};t-t_{i}) & \text{for $r_{1} < c(t-t_{i}) + r^{'}_{2}$}.
\end{cases}
\end{eqnarray}
Note that the dependence of the transition amplitudes for photon 2 on the output coordinate of photon 1 makes it impossible to separate the dynamics of photon 1 and photon 2. Therefore it is not possible to illustrate this time region using single photon amplitudes.
\end{description}

The matrix element with \(r^{'}_{1} < r^{'}_{2}\) can be obtained from the results for \(r^{'}_{2} < r^{'}_{1}\) by using the positive symmetry between photons propagating in free space. The results of these time regions can be summarized as follows,
\begin{eqnarray}
{\bf u}_{\text{2photon}}(r_{1},r_{2};r_{1}^{'},r_{2}^{'};t-t_{i}) &=& {\bf u}^{{\bf lin}}_{\text{2photon}}(r_{1},r_{2};r_{1}^{'},r_{2}^{'};t-t_{i}) + \Delta {\bf u}^{{\bf Nonlin}}_{\text{2photon}}(r_{1},r_{2};r_{1}^{'},r_{2}^{'};t-t_{i}) \nonumber \\
{\bf u}^{{\bf lin}}_{\text{2photon}}(r_{1},r_{2};r_{1}^{'},r_{2}^{'};t-t_{i}) &\equiv& {\bf u}_{\text{1photon}}(r_{1};r_{1}^{'};t-t_{i}) \cdot {\bf u}_{\text{1photon}}(r_{2};r_{2}^{'};t-t_{i}) \nonumber \\
\Delta {\bf u}^{{\bf Nonlin}}_{\text{2photon}}(r_{1},r_{2};r_{1}^{'},r_{2}^{'};t-t_{i}) &\equiv&
-\frac{4 \Gamma^{2}}{c^{2}} e^{-\frac{\Gamma}{c}(r_{1}^{'}+r_{2}^{'}+2c(t-t_{i})-r_{1}-r_{2})} \ \ \ \ \text{for} \ \ 0 < r_{1},r_{2} < c(t-t_{i}) + {\bf Min}[r_{1}^{'},r_{2}^{'}], \label{eq:twophotonpropagator}
\end{eqnarray}
where \({\bf Min}[r_{1}^{'},r_{2}^{'}]\) is the minimum of \(r^{'}_{1}\) and \(r^{'}_{2}\). The matrix element \(\Delta {\bf u}^{{\bf Nonlin}}_{\text{2photon}}\) of the nonlinear interaction between the two photons originates from the impossibility of two photon absorption at the single two-level atom. The remainder of the dynamics is identical to the single photon processes.
The total output wave function of two photons propagating in the one-dimensional field can then be expressed as
\begin{eqnarray}
\Psi(r_{1},r_{2};t) &=& \int^{\infty}_{-\infty} dr^{'}_{1}dr^{'}_{2} \ {\bf u}_{\text{2photon}}(r_{1},r_{2};r_{1}^{'},r_{2}^{'};t-t_{i}) \cdot \Psi(r^{'}_{1},r^{'}_{2};t_{i}). \label{eq:twophotonamplitude}
\end{eqnarray}
The output wave function describes the state of the photons propagating in the far field after the interaction with the atom. In general, a two photon wave function propagating in one-dimensional free space obeys the relation \(\Psi(r_{1},r_{2};t)=\Psi(r_{1}-ct,r_{2}-ct;0)\). Therefore the function of 3 parameters \(\Psi(r_{1},r_{2};t_{i})\) can be expressed by the function of 2 parameters \(\Psi_{\text{in}}(r_{1}-ct_{i},r_{2}-ct_{i})\). In the same way, the two photon wave function \(\Psi(r_{1},r_{2};t_{f})\) in the outgoing far field can be expressed as \(\Psi_{\text{out}}(r_{1}-ct_{f},r_{2}-ct_{f})\). We can then simplify our result (\ref{eq:twophotonamplitude}) by the transformation to a moving  coordinate system,
\begin{eqnarray}
r_{1}-ct &=& x_{1} \nonumber\\
r_{2}-ct &=& x_{2} 
\end{eqnarray}
In this coordinate system, the output state at an arbitrary time \(t_{f}\) can be expressed as 
\begin{eqnarray}
\Psi_{\text{out}}(x_{1},x_{2}) &=& \Psi(r_{1},r_{2};t_{f}) \nonumber \\
&=& \int^{\infty}_{-\infty} dx^{'}_{1}dx^{'}_{2} \ {\bf u}(x_{1},x_{2};x_{1}^{'},x_{2}^{'}) \cdot \Psi_{\text{in}}(x^{'}_{1},x^{'}_{2}) \nonumber\\
\text{with} && {} x_{1} = r_{1}-ct_{f}, \ \ x_{2} = r_{2}-ct_{f}, \nonumber\\
&& {} x^{'}_{1} = r^{'}_{1}-ct_{i}\ \ \text{and}\ \ x^{'}_{2} = r^{'}_{2}-ct_{i}, \label{eq:simplification}
\end{eqnarray}
where \({\bf u}(x_{1},x_{2};x_{1}^{'},x_{2}^{'})\) is given by
\begin{eqnarray}
{\bf u}(x_{1},x_{2};x_{1}^{'},x_{2}^{'}) &=& {\bf u}^{{\bf lin}}(x_{1},x_{2};x_{1}^{'},x_{2}^{'}) + \Delta {\bf u}^{{\bf Nonlin}}(x_{1},x_{2};x_{1}^{'},x_{2}^{'}), \label{eq:sipgenerator}\\
\text{where} && {} {\bf u}^{{\bf lin}}(x_{1},x_{2};x_{1}^{'},x_{2}^{'}) = {\bf u}_{\text{1photon}}(x_{1};x_{1}^{'}) \cdot {\bf u}_{\text{1photon}}(x_{2};x_{2}^{'}) \label{eq:twooutlin} \\
\text{with} && {} {\bf u}_{\text{1photon}}(x;x^{'}) = \delta(x-x^{'})-\frac{2\Gamma}{c} e^{-\frac{\Gamma}{c}(x^{'}-x)} \ \ \ \ \text{for $x \leq x^{'}$} \label{eq:oneoutgene}\\
\text{and} && {} \Delta {\bf u}^{{\bf Nonlin}}(x_{1},x_{2};x_{1}^{'},x_{2}^{'}) = -\frac{4 \Gamma^{2}}{c^{2}} e^{-\frac{\Gamma}{c}(x_{1}^{'}+x_{2}^{'}-x_{1}-x_{2})} \ \ \ \ \text{for} \ \ x_{1},x_{2} < {\bf Min}[x_{1}^{'},x_{2}^{'}]. \label{eq:nonlineffect}
\end{eqnarray}
Note that the linear component \({\bf u}^{{\bf lin}}(x_{1},x_{2};x^{'}_{1},x^{'}_{2})\) of the matrix element \({\bf u}(x_{1},x_{2};x_{1}^{'},x_{2}^{'})\) represents the time evolution of the two photons without interaction. Therefore it can be expressed by the product of two one photon matrix elements. On the other hand, the nonlinear component \(\Delta {\bf u}^{{\bf Nonlin}}(x_{1},x_{2};x_{1}^{'},x_{2}^{'})\) represents the difference from the time evolution of the linear component.  The output wave function \(\Psi_{\text{out}}(x_{1},x_{2})\) can then be expressed as the superposition of the linear term and the nonlinear term by substituting the right hand side of (\ref{eq:sipgenerator}) into (\ref{eq:simplification}),
\begin{eqnarray}
\Psi_{\text{out}}(x_{1},x_{2}) &=& \Psi^{{\bf lin}}_{\text{out}}(x_{1},x_{2}) + \Delta \Psi^{{\bf Nonlin}}_{\text{out}}(x_{1},x_{2}) \label{eq:resimplification}\\
\Psi^{{\bf lin}}_{\text{out}}(x_{1},x_{2}) &=& \int^{\infty}_{-\infty} dx^{'}_{1}dx^{'}_{2} \ {\bf u}^{{\bf lin}}(x_{1},x_{2};x_{1}^{'},x_{2}^{'}) \cdot \Psi_{\text{in}}(x^{'}_{1},x^{'}_{2}) \label{eq:linearterm}\\
\Delta \Psi^{{\bf Nonlin}}_{\text{out}}(x_{1},x_{2}) &=& \int^{\infty}_{-\infty} dx^{'}_{1}dx^{'}_{2} \ \Delta{\bf u}^{{\bf Nonlin}}(x_{1},x_{2};x_{1}^{'},x_{2}^{'}) \cdot \Psi_{\text{in}}(x^{'}_{1},x^{'}_{2}). \label{eq:nonlinearterm}
\end{eqnarray}
The formulation above can be used to analyze the outgoing amplitude for any arbitrary incoming two photon state. In the next section, these results are applied to the case of rectangular two photon input wave packets.
\section{\label{sec:level6}General solution for rectangular input wave packets}
In order to investigate the typical properties of the nonlinear photon-photon interaction, a rectangular two photon input wave packet is convenient because the homogeneous probability distribution of the two input photons makes it easier to understand the change of the correlations between the two photons due to the interaction. We assume that the shape of a single two photon pulse prepared in the incoming far field is a rectangle of \(L\) in length. Such a rectangular two photon wave packet can be written as follows,
\begin{eqnarray}
\Psi_{\text{in}}(x_{1},x_{2}) &=& \Psi_{\text{in}}(x_{1}) \cdot \Psi_{\text{in}}(x_{2}) \nonumber\\
\Psi_{\text{in}}(x) &=&
\begin{cases}
\sqrt{\frac{1}{L}} & \text{for $0 \le x \le L$}\\
0 & \text{else.}
\end{cases} \label{eq:oneoutput}
\end{eqnarray}
Note that, in any practical situation, the flanks of a rectangular pulse will not rise and fall infinitely fast. The discontinuities of the wavefunction should therefore be interpreted as a continuous change of amplitude that is extremely fast on a timescale of $1/\Gamma$. In particular, the flanks of the pulse should be smooth at a timescale of $1/\kappa$ due to the limitations of the model regarding the description of the cavity dynamics (see section \ref{sec:level3}). However, we assume this timescale to be so much shorter than $1/\Gamma$ that its effects can be neglected in the following.

\begin{figure}[ht]
\begin{minipage}{.45\linewidth}
\begin{picture}(0,0)
\put(-140,-8){(a)}
\end{picture}
\includegraphics[width=\linewidth,height=\linewidth]{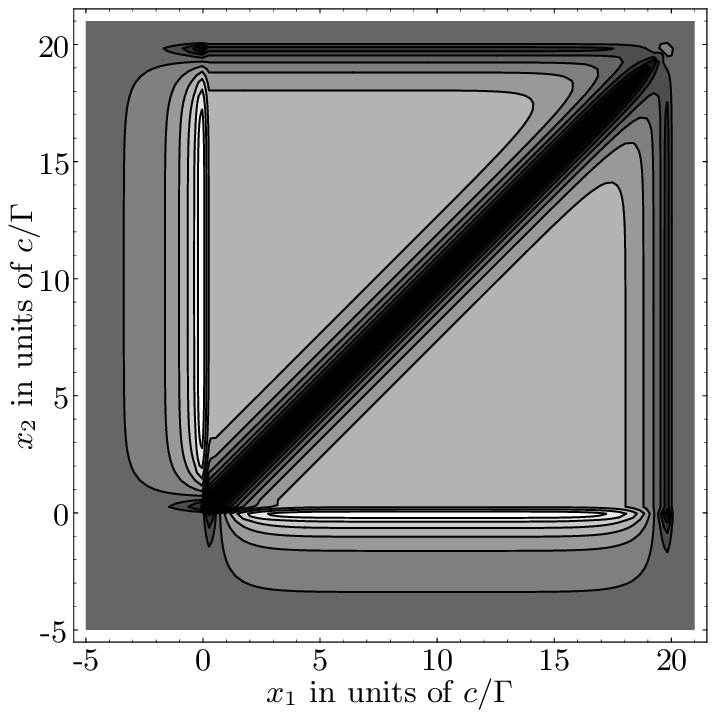}
\end{minipage}
\ \ \ \ \ \ \ \ \ \ \ \ 
\begin{minipage}{.45\linewidth}
\begin{picture}(0,0)
\put(-140,-8){(b)}
\end{picture}
\includegraphics[width=\linewidth,height=\linewidth]{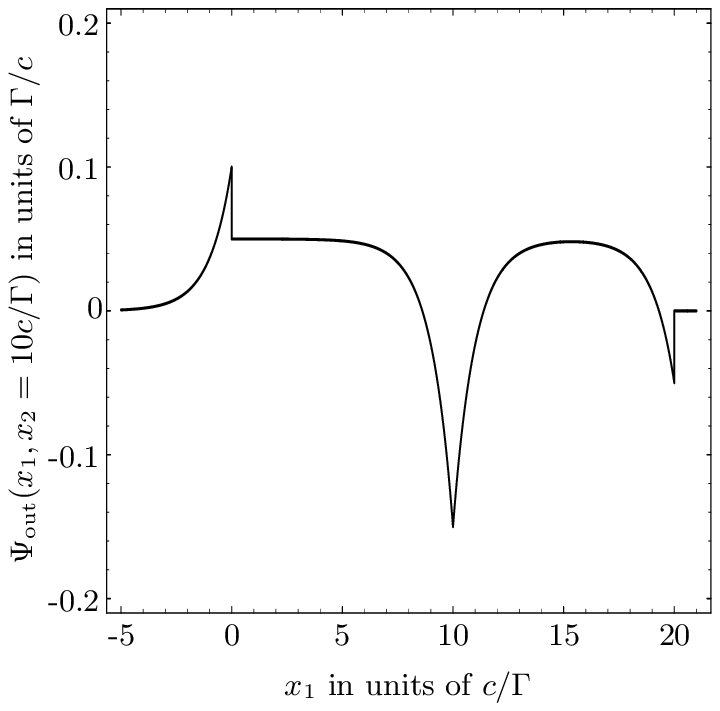}
\end{minipage}
\caption{\label{fig:figure5}Contour plot of the output two photon wave packet for an input wave packet length of \(L=20c/\Gamma\). The horizontal axes \(x_{1}\) in both figures represent the space coordinate of photon 1, and the vertical axes \(x_{2}\) represent the space coordinate of photon 2.  (a) is a contour line plot of \(\Psi_{\text{out}}(x_{1},x_{2})\) and (b) is the cross section of the contour plot at \(x_{2}=10c/\Gamma\)}
\end{figure}

As shown in (\ref{eq:resimplification}), the output wave packet can be separated into a linear term and a nonlinear term. The linear term in the output of the rectangular input wave packet described by eq.(\ref{eq:oneoutput}) can be obtained according to eq.(\ref{eq:linearterm}). It reads 
\begin{eqnarray}
\Psi^{{\bf lin}}_{\text{out}}(x_{1},x_{2}) &=& \int^{\infty}_{-\infty} dx^{'}_{1}dx^{'}_{2} \ {\bf u}^{{\bf lin}}(x_{1},x_{2};x_{1}^{'},x_{2}^{'}) \cdot \Psi_{\text{in}}(x^{'}_{1},x^{'}_{2}) = \Psi_{\text{out}}(x_{1}) \cdot \Psi_{\text{out}}(x_{2}) \nonumber\\
\text{with} && {} \Psi_{\text{out}}(x) = \int^{\infty}_{- \infty} dx^{'} \ u_{\text{1photon}}(x;x^{'}) \cdot \Psi_{\text{in}}(x^{'}) \nonumber \\
&& {}\ \ \ \ \ \ \ \ \ \ =
\begin{cases}
\frac{2}{\sqrt{L}} \left(e^{-\frac{\Gamma}{c} (L-x)}-e^{\frac{\Gamma}{c} x} \right) & \text{for $x < 0$}\\
\frac{1}{\sqrt{L}} \left(2 e^{-\frac{\Gamma}{c} (L-x)}-1 \right) & \text{for $0 \le x \le L$}\\
0 & \text{else.}
\end{cases} \label{eq:reconephoton}
\end{eqnarray}
The nonlinear term in the output of the rectangular input wave packet described by eq.(\ref{eq:oneoutput}) can be obtained according to eq.(\ref{eq:nonlinearterm}). It reads
\begin{eqnarray}
\Delta \Psi^{{\bf Nonlin}}_{\text{out}}(x_{1},x_{2}) &=& \int^{\infty}_{-\infty} dx^{'}_{1}dx^{'}_{2} \ \Delta {\bf u}^{{\bf Nonlin}}(x_{1},x_{2};x_{1}^{'},x_{2}^{'}) \cdot \Psi_{\text{in}}(x^{'}_{1},x^{'}_{2}) \nonumber \\
&=& -\frac{4}{L}e^{\frac{\Gamma}{c}(x_{1}+x_{2})} \left( e^{-\frac{\Gamma}{c} {\bf Max} [0,x_{1},x_{2}]}-e^{-\frac{\Gamma}{c}L} \right)^{2} \ \ \ \ \text{for $x_{i} \le L$ $(i=1,2)$}, \label{eq:recnonlineffect}
\end{eqnarray}
where \({\bf Max} [0, x_{1}, x_{2}]\) is the maximum of \(0,x_{1}\) and \(x_{2}\). The output two photon wave packet can then be written as
\begin{eqnarray}
\Psi_{\text{out}}(x_{1},x_{2}) &=& \Psi^{\bf{lin}}_{\text{out}}(x_{1},x_{2}) +\Delta \Psi^{{\bf Nonlin}}_{\text{out}}(x_{1},x_{2}). \label{eq:rectangular}
\end{eqnarray}
It should be noted that the shape of the output wave packet is defined by the ratio of the dipole relaxation length \(c/\Gamma\) and the length \(L\) of the input wave packet. In frequency representation, this means that the shape of the output wave packet is sensitive to whether the frequency spectrum of the input wave packet is narrower than the atomic line width \(2\Gamma\) or not. A particularly simple case for showing the contribution of the nonlinear term to the output wave packet can be obtained in the long pulse limit \(c/\Gamma \ll L\) because, in this limit, the linear term given by eq.(\ref{eq:reconephoton}) of the output wave packet is almost constant for the region \(0 < x_{i} < L-2c/\Gamma\ (i=1,2)\), that is, 
\begin{eqnarray}
\Psi^{{\bf lin}}_{\text{out}}(x_{1},x_{2}) &\simeq& \frac{1}{L} \ \ \ \ \text{for $0 < x_{i} < L-2c/\Gamma$ $(i=1,2)$.}
\end{eqnarray}
On the other hand, the nonlinear term given by eq.(\ref{eq:recnonlineffect}) of the output wave packet can be approximated as
\begin{eqnarray}
\Delta \Psi^{{\bf Nonlin}}_{\text{out}}(x_{1},x_{2}) &\simeq& -\frac{4}{L} e^{-\frac{\Gamma}{c} \left|x_{1}-x_{2}\right|} \ \ \ \ \text{for $0 < x_{i} < L-2c/\Gamma$ $(i=1,2)$}
\end{eqnarray}
which only depends on the relative distance \(\left|x_{1}-x_{2}\right|\) between the two photons. The exponential decay indicates that the nonlinear deviation from the linear term only becomes significant in the vicinity of \(x_{1}=x_{2}\). The output wave function can then be written as
\begin{eqnarray}
\Psi_{\text{out}}(x_{1},x_{2}) &\simeq& \frac{1}{L}-\frac{4}{L} e^{-\frac{\Gamma}{c} \left|x_{1}-x_{2}\right|} \ \ \ \ \text{for $0 < x_{i} < L-2c/\Gamma$ $(i=1,2)$.} \label{eq:longpulselimit}
\end{eqnarray}
\ \ The contour plot in fig.~\ref{fig:figure5} (a) shows an example of the output wave function \(\Psi_{\text{out}}(x_{1},x_{2})\) in the long pulse limit. The probability amplitude increases from black to white shading. In this example, we have chosen an input wave packet length \(L=20c/\Gamma\) which is 20 times greater than dipole relaxation length \(c/\Gamma\). The plateau region away from \(x_{1}=x_{2}\) in  fig.~\ref{fig:figure5} (a) has positive amplitude. On the other hand, in the vicinity of \(x_{1}=x_{2}\), a valley of negative amplitude cuts across this plateau. The shape of this valley can be seen in  fig.~\ref{fig:figure5} (b).  Fig.~\ref{fig:figure5} (b) is the cross-section of the contour plot at \(x_{2}=10c/\Gamma\). It should be noted that the shape of the valley is the same for any cross-section \(x_{2}\) within the plateau region. 
The valley is due to the contribution of the nonlinear term in eq.(\ref{eq:recnonlineffect}) which decreases with the distance between \(x_{1}\) and \(x_{2}\). The plateau is the unchanged characteristic feature of the long rectangular input wave packet. Fig.\ref{fig:figure5} shows the typical characteristics in the output for a long two photon input pulse. In the next section, these typical characteristics are analyzed in terms of two photon statistics.
\section{\label{sec:level7}Analysis of two photon statistics}
\begin{figure}[ht]
\begin{minipage}{.40\linewidth}
\begin{picture}(0,0)
\put(-130,-8){(a-1)}
\end{picture}
\includegraphics[width=\linewidth,height=\linewidth]{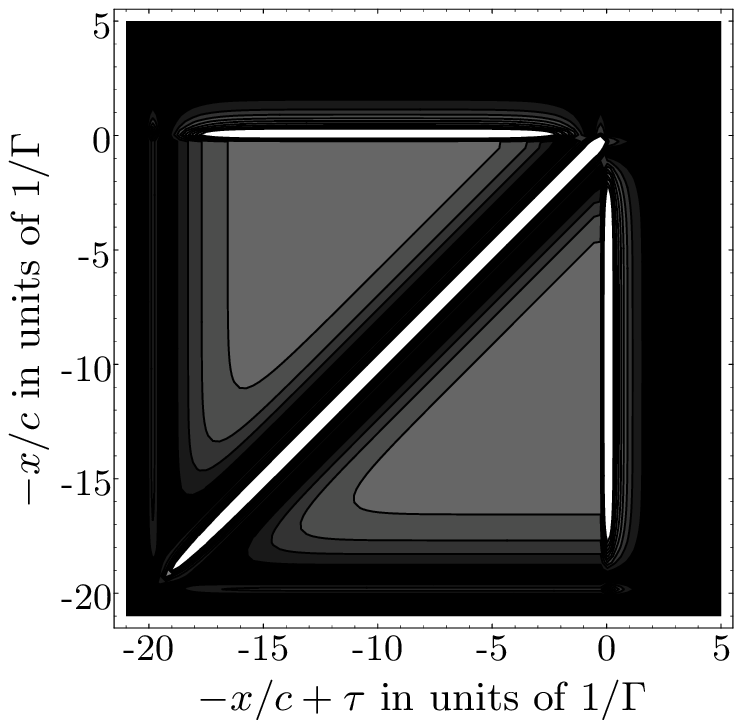}
\end{minipage}
\ \ \ \ \ \ \ \ \ \ \ \ \ 
\begin{minipage}{.40\linewidth}
\begin{picture}(0,0)
\put(-130,-8){(a-2)}
\end{picture}
\includegraphics[width=\linewidth,height=\linewidth]{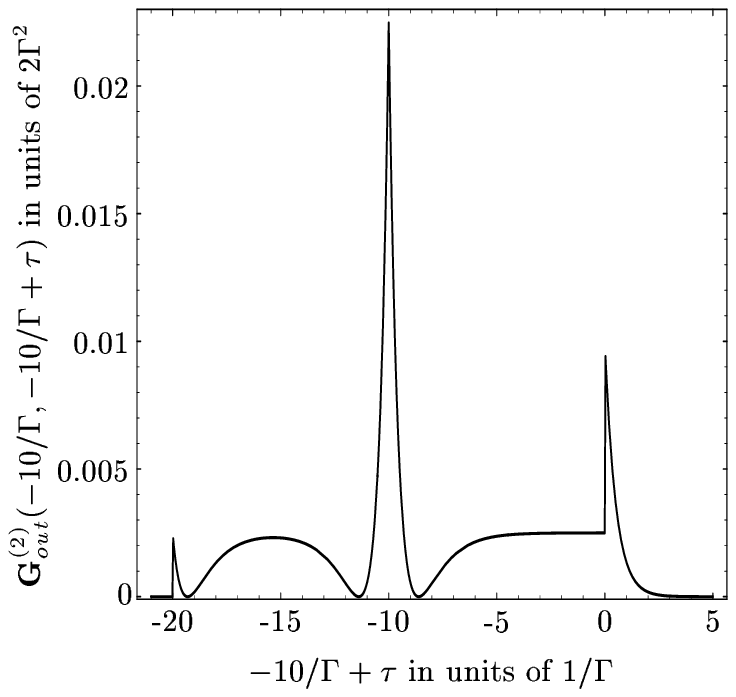}
\end{minipage}
\begin{minipage}{.40\linewidth}
\vspace*{1cm}
\begin{picture}(0,0)
\put(-130,-8){(b-1)}
\end{picture}
\includegraphics[width=\linewidth,height=\linewidth]{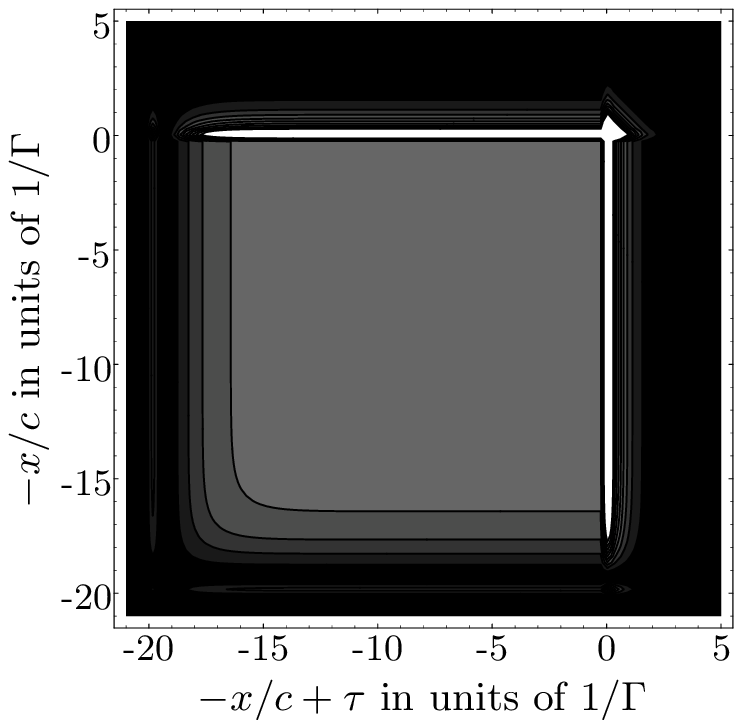}
\end{minipage}
\ \ \ \ \ \ \ \ \ \ \ \ \ 
\begin{minipage}{.40\linewidth}
\vspace*{1cm}
\begin{picture}(0,0)
\put(-130,-8){(b-2)}
\end{picture}
\includegraphics[width=\linewidth,height=\linewidth]{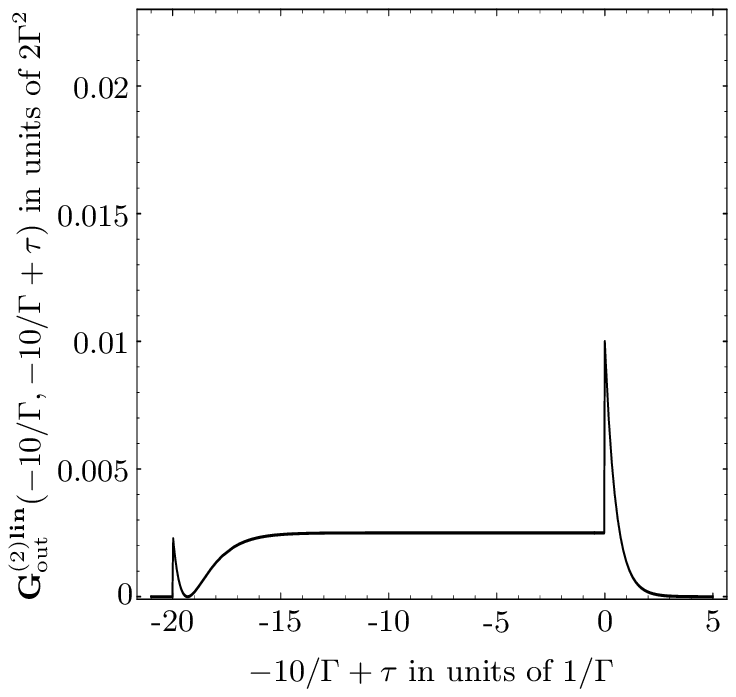}
\end{minipage}
\caption{\label{fig:figure6}Distribution of the second order correlations (a) \({\bf G}^{(2)}_{\text{out}}(-x/c,-x/c+\tau)\) and (b) \({\bf G}^{(2){\bf lin}}_{\text{out}}(-x/c+\tau,-x/c)\) for an input wave packet length of \(L=20c/\Gamma\). The horizontal axes \(-x/c+\tau\) in (a-1) and (b-1) represent the detection time for one photon, and the vertical axes \(-x/c\) represent the detection time for the other photon. The second order correlations increase from black to white shading. (a-2) and (b-2) are the cross-sections of (a-1) and (b-1) at \(t=10/\Gamma\).}
\end{figure}

Two photon statistics are often used to characterize the properties of nonclassical light from the viewpoint of photon counting. For this purpose, the second order correlation \({\bf G}^{(2)}_{\text{out}}(t,t+\tau)\) is defined by the joint probability density of detecting one photon at time \(t+ \tau\) when the other photon is detected at time \(t\). When the second order correlation \({\bf G}^{(2)}_{\text{out}}(t,t+\tau)\) is expressed by using the probability density \(\left|\Psi_{\text{out}}(x+c\tau,x)\right|^{2}\), the time t is given by the space coordinate x divided by the speed of light, \(t=-x/c\). Also, the probability density \(\left|\Psi_{\text{out}}(x+c\tau,x)\right|^{2}\) must be rescaled by a factor of \(c^{2}\), since the probability density per unit time is expressed by c times the probability density per unit length and, for second order correlations, this factor is squared. Moreover \({\bf G}^{(2)}_{\text{out}}\) does not distinguish between the two photons. Therefore, both \(\left|\Psi_{\text{out}}(x,x+c\tau)\right|^{2}\) and \(\left|\Psi_{\text{out}}(x+c\tau,x)\right|^{2}\) contribute to \({\bf G}^{(2)}_{\text{out}}(t,t+\tau)\). Taking all these factors into account, the second order correlation function can be written as 
\begin{eqnarray}
{\bf G}^{(2)}_{\text{out}}(-x/c,-x/c+\tau) &=& 
c^{2} \left( \left| \Psi_{\text{out}}(x+c\tau,x) \right|^{2} + \left| \Psi_{\text{out}}(x,x+c\tau) \right|^{2} \right) = 2c^{2}\left| \Psi_{\text{out}}(x+c\tau,x) \right|^{2}. \label{eq:second}
\end{eqnarray}
Note that \(\left| \Psi_{\text{out}}(x+c\tau,x) \right|^{2}\) and  \(\left| \Psi_{\text{out}}(x,x+c\tau) \right|^{2}\) always have to be equal to each other because of the bosonic nature of the two photons. The contour plot of the second order correlations is shown in fig.~\ref{fig:figure6}(a-1) and (a-2) for an input pulse length of \(L=20c/\Gamma\). The correlation increases from black to white shading. To indicate the contribution of the nonlinear term described by eq.(\ref{eq:recnonlineffect}) in the second order correlations, it is convenient to compare this result with the second order correlations obtained from the linear component of the output only. Fig.~\ref{fig:figure6} (b-1) and (b-2) shows this second order correlation \({\bf G}^{(2){\bf lin}}_{\text{out}}(-x/c,-x/c+\tau) = 2c^{2} \left| \Psi^{{\bf lin}}_{\text{out}}(x+c\tau,x) \right|^{2}\) of the linear component. The comparison of these figures shows that both cases are identical except for their distributions around \(\tau=0\). As can be seen in fig.~\ref{fig:figure6} (a-2), the nonlinear interaction causes photon bunching around \(\tau =0\) and photon anti-bunching around \(\tau = \pm 2 \log 2/\Gamma\).

\begin{figure}[ht]
\begin{center}
\includegraphics[width=.45\linewidth]{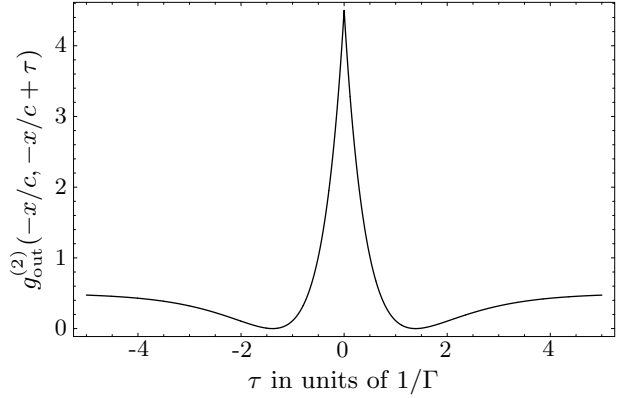}
\end{center}
\caption{\label{fig:figure7}Normalized second order correlation  \(g^{(2)}_{\text{out}}\) in the outgoing two photon wave packet. A typical feature of the correlations is the double dip feature with zero two-photon coincidence occuring at nonzero time delays \(\pm 2 \log 2/\Gamma\). \(g^{(2)}_{\text{out}}\) approaches \(1/2\) beyond the double dip feature, since \(g^{(2)}_{\text{out}}=1/2\) is a statistical property of the single mode two photon input state.}
\end{figure}

In order to compare the second order correlations with other systems in quantum optics, it is useful to normalize the correlation function \({\bf G}^{(2)}_{\text{out}}\) by the product of the probability densities of single photon detection at times \(-x/c\) and \(-x/c+\tau\) in the output field. In the long pulse limit, the total probability of finding both photons within the output region of \(0 < x_{i} < L\ (i=1,2)\) is nearly equal to \(1\). Therefore the single photon detection probability density per unit time within this region is \(2c/L\). By using this average photon density, we obtain the normalized second order correlation function,
\begin{eqnarray}
g^{(2)}_{\text{out}}(-x/c,-x/c+\tau) &=& {\bf G}^{(2)}_{\text{out}}(-x/c,-x/c+\tau)/(2c/L)^{2} \nonumber\\
&=& \frac{L^{2}}{2} \left| \Psi_{\text{out}}(x+c\tau,x) \right|^{2}.
\end{eqnarray}
With the approximations for the long pulse limit given by eq.(\ref{eq:longpulselimit}), this correlation reads
\begin{eqnarray}
g^{(2)}_{\text{out}}(-x/c,-x/c+\tau) &\simeq& \frac{1}{2} \left(1-4e^{-\Gamma \left| \tau \right|}\right)^{2} \ \ \ \ \text{for $0 \lesssim \{x+c\tau,x\} \lesssim L-2c/\Gamma$.}
\end{eqnarray}
That is, the second order correlation \(g^{(2)}_{\text{out}}(-x/c,-x/c + \tau)\) in the long pulse limit only depends on the delay time \(\tau\) in the output region \(0 \lesssim \{x,x+c\tau\} \lesssim L-2c/\Gamma\). \(g^{(2)}_{\text{out}}\) around \(\tau = 0\) is shown in fig.~\ref{fig:figure7}. A typical feature of the correlations is the double dip feature with zero two-photon coincidence occuring at nonzero time delays of \(\tau = \pm 2 \log 2/\Gamma\). Such a nonlinear effect is similar to the nonclassical effect which has been reported in \cite{rice,lu}. However, in our case, \(g^{(2)}_{\text{out}}\) approaches \(1/2\) beyond the double dip feature. The value \(g^{(2)}_{\text{out}}=1/2\) is a statistical property of the single mode two photon input state. Note that the value of \(g^{(2)}_{\text{out}}=1/2\) does not indicate that there is a correlation between the two photons in the pulse. The anti-correlation expressed by \(g^{(2)}_{\text{out}}=1/2\) simply arises because one cannot detect the same photon twice. In the next section, we trace the second order correlation back to the coherent terms of the output wave packet.
\section{\label{sec:level8}Explanation of the second order correlation function in terms of quantum interference effects}
\begin{figure}[ht]
\begin{minipage}{.40\linewidth}
\begin{picture}(0,0)
\put(-130,-10){(a)}
\end{picture}
\includegraphics[width=\linewidth,height=\linewidth]{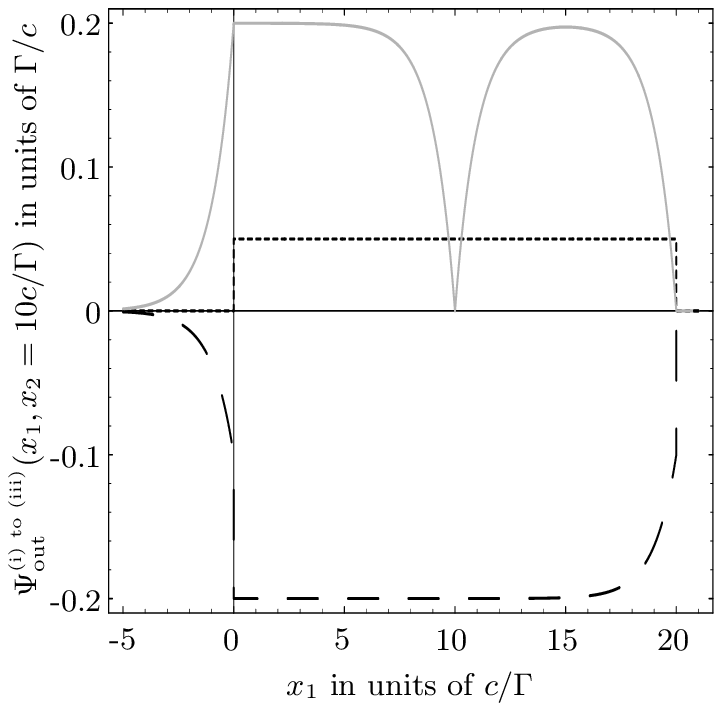}
\end{minipage}
\ \ \ \ \ \ \ \ \ \ \ \ \  
\begin{minipage}{.40\linewidth}
\begin{picture}(0,0)
\put(-130,-10){(b)}
\end{picture}
\includegraphics[width=\linewidth,height=\linewidth]{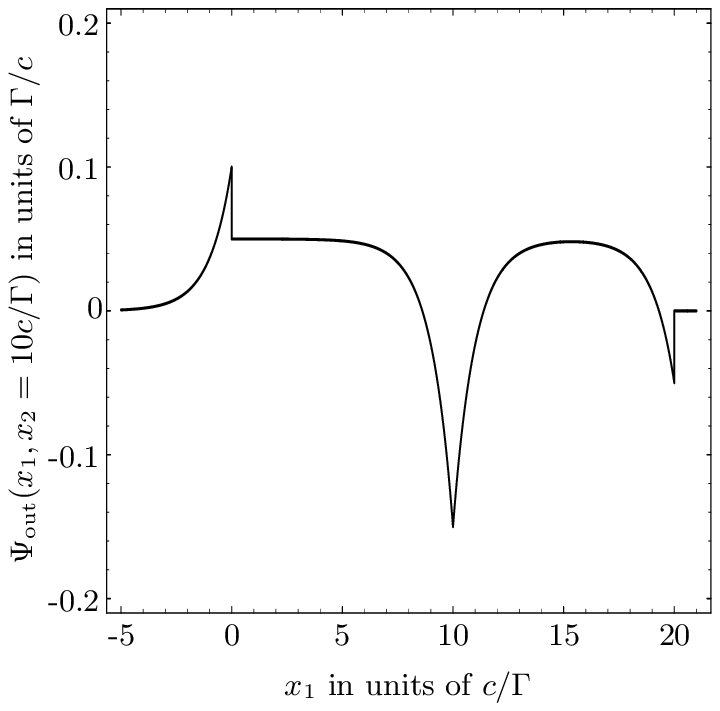}
\end{minipage}
\caption{\label{fig:figure8}Analysis of the interaction processes of two photons and the atom for an input pulse length of \(L=20c/\Gamma\). In (a), the dotted line shows the amplitude  at \(x_{2} = 10 c/\Gamma\) corresponding to process (i), where both photons are transmitted without absorption. The short dashed line shows the amplitude corresponding to process (ii), where one photon is absorbed and then reemitted while the another photon is transmitted without absorption. The thin line shows the amplitude corresponding to process (iii), where both photons are absorbed and then reemitted. The superposition of all the amplitudes is shown in (b). Note that (b) is identical with fig.~\ref{fig:figure5} (b).}
\end{figure}

The cause of the features of the second order correlation shown in fig.~\ref{fig:figure6} (a-2) and fig.~\ref{fig:figure7} can be understood by analyzing the interaction processes of two photons and the atom as follows. As shown in eq.(\ref{eq:onephotonprocesses}), the matrix element of the time evolution \({\bf u}_{{\bf 1photon}}(x;x^{'})\) can be expanded in terms of two interaction processes: (I) single photon transmission (reflection in fig.\ref{fig:figure2}) without absorption; (II) single photon reemission after absorption. In the same way, the matrix element of the time evolution \({\bf u}(x_{1},x_{2};x_{1}^{'},x_{2}^{'})\) can be expanded in terms of three interaction processes: (i) two photon transmission without absorption; (ii) one photon transmission without absorption and one photon reemission after absorption; (iii) two photon reemission after absorption,
\begin{eqnarray}
{\bf u}(x_{1},x_{2};x_{1}^{'},x_{2}^{'}) &=& {\bf u}^{\text{(i)}}(x_{1},x_{2};x_{1}^{'},x_{2}^{'}) + {\bf u}^{\text{(ii)}}(x_{1},x_{2};x_{1}^{'},x_{2}^{'}) + {\bf u}^{\text{(iii)}}(x_{1},x_{2};x_{1}^{'},x_{2}^{'}).
\end{eqnarray}
For \(x^{'}_{1} \le x_{1}\) and \(x^{'}_{2} \le x_{2}\), the components of \({\bf u}(x_{1},x_{2};x_{1}^{'},x_{2}^{'})\) read
\begin{eqnarray}
{\bf u}^{\text{(i)}}(x_{1},x_{2};x_{1}^{'},x_{2}^{'}) &=& {\bf u}_{\text{prop}}(x_{1};x^{'}_{1}) \cdot {\bf u}_{\text{prop}}(x_{2};x^{'}_{2}) \nonumber\\
{\bf u}^{\text{(ii)}}(x_{1},x_{2};x_{1}^{'},x_{2}^{'}) &=& {\bf u}_{\text{prop}}(x_{1};x^{'}_{1}) \cdot {\bf u}_{\text{abs}}(x_{2};x^{'}_{2})  +  {\bf u}_{\text{abs}}(x_{1};x^{'}_{1}) \cdot {\bf u}_{\text{prop}}(x_{2};x^{'}_{2})\nonumber\\
{\bf u}^{\text{(iii)}}(x_{1},x_{2};x_{1}^{'},x_{2}^{'}) &=& {\bf u}_{\text{abs}}(x_{1};x^{'}_{1}) \cdot {\bf u}_{\text{abs}}(x_{2};x^{'}_{2}) + \Delta {\bf u}^{{\bf Nonlin}}(x_{1},x_{2};x_{1}^{'},x_{2}^{'}).
\end{eqnarray}
Therefore, the output wave packet \(\Psi_{\text{out}}(x_{1},x_{2})\) can be expanded as\begin{eqnarray}
\Psi_{\text{out}}(x_{1},x_{2}) &=& \Psi^{\text{(i)}}_{\text{out}}(x_{1},x_{2}) + \Psi^{\text{(ii)}}_{\text{out}}(x_{1},x_{2}) + \Psi^{\text{(iii)}}_{\text{out}}(x_{1},x_{2}), \nonumber
\end{eqnarray}
where \(\Psi^{\text{(i) to (iii)}}\) is given by 
\begin{eqnarray}
\Psi_{\text{out}}^{\text{(i) to (iii)}}(x_{1},x_{2})  &=& \int^{\infty}_{-\infty} dx^{'}_{1}dx^{'}_{2} {\bf u}^{\text{(i) to (iii)}}(x_{1},x_{2},x_{1}^{'},x_{2}^{'}) \cdot \Psi_{\text{in}}(x^{'}_{1},x^{'}_{2}).
\end{eqnarray}
For the rectangular input wave packet, the output wave function in the interval \(0 \le x_{i} \le L\ (i=1,2)\) is given by
\begin{eqnarray}
\Psi^{\text{(i)}}_{\text{out}}(x_{1},x_{2}) &=& \frac{1}{L} \nonumber \\
\Psi^{\text{(ii)}}_{\text{out}}(x_{1},x_{2}) &=& \frac{1}{L}\left(2 e^{-\frac{\Gamma}{c}(L-x_{1})}-2\right) + \frac{1}{L}\left(2 e^{-\frac{\Gamma}{c}(L-x_{2})}-2\right)  \nonumber \\
\Psi^{\text{(iii)}}_{\text{out}}(x_{1},x_{2}) &=& \frac{4}{L} \left(e^{-\frac{\Gamma}{c}(L-x_{1})}-1\right)\left(e^{-\frac{\Gamma}{c}(L-x_{2})}-1\right) + \Delta \Psi^{{\bf Nonlin}}(x_{1},x_{2}). \label{eq:3processes}
\end{eqnarray}
Fig.~\ref{fig:figure8} (a) shows the correlations of \(\Psi^{\text{(i) to (iii)}}_{\text{out}}(x_{1},x_{2})\) at \(x_{2}=10c/\Gamma\) for an input pulse length of \(L=20c/\Gamma\). The dotted line shows the amplitude corresponding to process (i), where both of two photons are transmitted without absorption by the atom. Likewise, the short dashed line shows the amplitude corresponding to process (ii), where one photon is absorbed and then reemitted and another photon is transmitted without absorption. The thin line shows the amplitude corresponding to process (iii), where both photons are absorbed and then reemitted. Fig.~\ref{fig:figure8} (b) shows the superposition of all the amplitudes. This superposition is equal to the output amplitude \(\Psi_{\text{out}}(x_{1},x_{2}=10c/\Gamma)\). Therefore fig.~\ref{fig:figure8} (b) is identical with fig.~\ref{fig:figure5} (b). Note that the discontinuities of the output wavefunction $\Psi_{\text{out}}$ at $x_{i}=0$ and $x_{i}=20c/\Gamma$ originate from the direct transmission of the rectangular input wavepacket in $\Psi_{\text{out}}^{\text{i}}$ and $\Psi_{\text{out}}^{\text{ii}}$. As mentioned in section \ref{sec:level6}, these discontinuities represent changes of the amplitude that are extremely fast on a timescale of $1/\Gamma$, but would be continuous at a much shorter timescale of $1/\kappa$. They therefore approximately represent the linear response of the atom-cavity system at time resolutions sufficiently longer than the timescale $1/\kappa$ of the cavity dynamics.

In the long pulse limit \(L \gg c/\Gamma\), the amplitudes described by eq.(\ref{eq:3processes}) can be approximated as
\begin{eqnarray}
\Psi^{\text{(i)}}_{\text{out}}(x_{1},x_{2}) &=& \frac{1}{L} \nonumber \\
\Psi^{\text{(ii)}}_{\text{out}}(x_{1},x_{2}) &\simeq& -\frac{4}{L} \nonumber \\
\Psi^{\text{(iii)}}_{\text{out}}(x_{1},x_{2}) &\simeq& \frac{4}{L} + \Delta \Psi^{{\bf Nonlin}}(x_{1},x_{2}), \nonumber\\ 
&& {} \text{where $\Delta \Psi^{{\bf Nonlin}}(x_{1},x_{2}) = -\frac{4}{L}e^{-\frac{\Gamma}{c}\left|x_{1}-x_{2}\right|}$}. \label{eq:longpulseprocesses}
\end{eqnarray}
To understand the details of the two time correlations, we now examine the effect of the non-linear contribution \(\Delta \Psi^{{\bf Nonlin}}_{\text{out}}(x_{1},x_{2})\) in the superposition. The non-linear contribution only depends on the relative distance \(\left|x_{2}-x_{1}\right|\) which corresponds to the difference between the detection times of the two photons. When the relative distance \(\left|x_{2}-x_{1}\right|\) is much larger than the dipole relaxation length \(c/\Gamma\), the non-linear contribution \(\Delta \Psi^{{\bf Nonlin}}_{\text{out}}\) is nearly equal to zero and the double absorption amplitude \(\Psi^{\text{(iii)}}_{\text{out}}\) is close to \(4/L\). On the other hand, when the relative distance \(\left|x_{2}-x_{1}\right|\) is much smaller than the dipole relaxation length \(c/\Gamma\), the non-linear contribution \(\Delta \Psi^{{\bf Nonlin}}_{\text{out}}\) is nearly equal to \(-4/L\) and the double absorption amplitude \(\Psi^{\text{(iii)}}_{\text{out}}\) is close to zero. As noted previously, \(\Delta \Psi^{{\bf Nonlin}}_{\text{out}}\) represents the absence of simultaneous double absorption. The time dependence of \(\Delta \Psi^{{\bf Nonlin}}_{\text{out}}\) therefore describes the saturation dynamics of the two-level atom. Since the no-absorption amplitude \(\Psi^{\text{(i)}}_{\text{out}}\) and the single absorption amplitude \(\Psi^{\text{(ii)}}_{\text{out}}\) are independent of the saturation and their amplitudes are \(1/L\) and \(-4/L\) respectively, the total amplitude \(\Psi_{\text{out}}\) is obtained by adding a constant value of \(-3/L\) to \(\Psi^{\text{(iii)}}_{\text{out}}\). The resulting total amplitude \(\Psi_{\text{out}}\) is then close to \(1/L\) for relative distances \(\left|x_{2}-x_{1}\right| \gg c/\Gamma\) and drops to \(-3/L\) for relative distances \(\left|x_{2}-x_{1}\right| \ll c/\Gamma\). The total amplitude \(-3/L\) for relative distances \(\left|x_{2}-x_{1}\right| \ll c/\Gamma\) is associated with the bunching effect in the second order correlations. The total amplitude then changes from positive to negative values continuously, depending only on the relative distance between the two photons \(\left|x_{2}-x_{1}\right|\). Precisely speaking, the total amplitude \(\Psi_{\text{out}}\) is positive if the relative distance is \(\left|x_{2}-x_{1}\right| > 2\log(2) c/\Gamma\) and negative if the relative distance is \(\left|x_{2}-x_{1}\right| < 2\log(2) c/\Gamma\) (see fig.~\ref{fig:figure8} (b)). That is, in the region with \(\left|x_{2}-x_{1}\right| < 2\log(2) c/\Gamma\), the interaction of the two photons causes a phase flip of \(\pi\). This phase flip can be understood as evidence on the quantum level for the resonant nonlinearity discussed in \cite{holkoji}. The anti-bunching at \(\left|x_{2}-x_{1}\right| = 2\log(2) c/\Gamma\) originates from \(\Psi_{\text{out}}\) passing through zero as the sign of the total amplitude changes from negative to positive. In this way, both the bunching and the anti-bunching effects in the two photon correlation can be explained in terms of interference effects between the quantum coherence contributions from the different interaction processes.\\
\section{\label{sec:level9}Conclusions}
We have presented a fully quantum mechanical model of the non-linear interaction of two photons at a two-level atom. The experimental realization of such an interaction can be implemented using a one-sided cavity and single photon sources. Our theory allows us to determine the effects of an atomic nonlinearity on the spatiotemporal coherence of a two photon state. By applying the general results to rectangular input wave packets, we have shown that the nonlinear interaction of two photons at the atom causes bunching and anti-bunching effects in the two photon output state. Since our model describes the complete spatiotemporal coherence of the output, it is possible to analyze the origin of these effects in terms of quantum interference between different absorption and propagation processes. This method may therefore provide a useful tool for various applications in the manipulation of individual photons such as quantum information processing, quantum nondemolition measurements, and entangled photon sources.
\begin{acknowledgments}
We wish to express our gratitude to M. Ozawa, K. Sakoda, A. Furusawa, and M. Kitano for helpful discussions. K.K. also wishes to thank H. Nishimura for his encouragement. This work was partly supported by the program "Research and Development on Quantum Communication Technology" of the Ministry of Public Management, 
Home Affairs, Posts and Telecommunications of Japan.
\end{acknowledgments}

\end{document}